\documentclass[preprintnumbers,secnumarabic,onecolumn]{revtex4}
\pdfoutput=1
\usepackage{graphicx}
\usepackage{amsmath,amssymb,mathrsfs,bm}
\usepackage{array}
\newcommand*{\ewgroup}{\ensuremath{SU(2)_L \times U(1)_Y}}
\newcommand*{\trans}{\mathrm{T}}                     % transposed
\newcommand*{\unitmatrix}{\bm{1}}
\newcommand*{\tvec}[1]{\ensuremath{\boldsymbol{\mathrm{#1}}}}           % 3 vec
\newcommand*{\tmat}[1]{\underline{#1}}             % 2x2 matrix

\makeatletter       % Macros for arabic section numbering

\renewcommand{\p@subsection}{}

\makeatother

\DeclareMathOperator{\re}{Re}
\DeclareMathOperator{\im}{Im}

\begin{document}

\author{Ernest Ma$^a$}
\author{Markos Maniatis$^b$}

\affiliation{a) Department of Physics and Astronomy, University of California,
Riverside, California 92521, USA}
\affiliation{b) Institut f\"ur Theoretische Physik, University of Heidelberg,
69120 Heidelberg, Germany}

\preprint{UCRHEP-T474, HD-THEP-09-17}

\title{Symbiotic Symmetries of the Two-Higgs-Doublet Model}

\begin{abstract}
The new phenomenon of {\em symbiotic symmetries} is described in the 
context of the Two-Higgs-Doublet Model (THDM).  The quartic potential 
has two or more separate sectors with unequal symmetries, but these 
unequal symmetries persist even though the different sectors are 
renormalized by one another.  We discuss all such symmetries of the THDM, 
consistent with the $SU(2) \times U(1)$ gauge interactions, using the Pauli 
formalism.
\end{abstract}

\maketitle

Much attention has been paid to the Two-Higgs-Doublet Model (THDM) which 
has a long history, with applications to many diverse issues in high-energy 
physics~\cite{Lee:1973iz,Kobayashi:1973fv,Gunion:1989we,Cvetic:1993cy,
Ginzburg:2004vp,Gunion:2005ja,Branco:2005em,Nishi:2006tg,Fromme:2006cm,
Maniatis:2006fs,Barroso:2007rr,Gerard:2007kn,Mahmoudi:2009zx,Nagel:2004sw,
Deshpande:1977rw,Maniatis:2007vn,Nishi:2007dv,Ferreira:2009wh,Ferreira:2009jb,Ma:2006km}. 
One important reason is supersymmetry, where the minimal extension of the 
Standard Model (SM) of quarks and leptons requires two Higgs doublet 
superfields.  On the other hand, even without supersymmetry, THDM's have 
interesting properties in their own right.  For example, CP violation is 
allowed in the Higgs sector itself, and there are five physical Higgs bosons 
to be searched for at the upcoming Large Hadron Collider (LHC), among many 
other possible considerations.\\

The most general THDM Higgs potential, which obeys electroweak gauge symmetry 
and is renormalizable, has three quadratic mass terms and seven quartic 
coupling terms. Experimentally, the non-observation of flavor-changing 
neutral currents limits the available parameter space, and may correspond 
to a symmetry~\cite{Glashow:1976nt,Paschos:1976ay,Fayet:1974pd,Peccei:1977hh,
Peccei:1977ur}.  If some version of the THDM is realized in Nature, 
it will be important to understand why certain terms of the general THDM 
should be suppressed.  An underlying symmetry principle may be responsible 
for this selection, which must of course be also phenomenologically 
acceptable.\\

In this work we look systematically for symmetries~\cite{Deshpande:1977rw,
Ma:1977td} in the most general THDM Higgs potential which are preserved by 
the renormalization-group equations (rge's) in the presence of gauge 
interactions.  We make use of the powerful new formalism recently proposed 
to describe the general THDM in a concise way~\cite{Nagel:2004sw,
Maniatis:2006fs,Nishi:2006tg,Maniatis:2007vn}.  In this formalism, all 
gauge-invariant expressions are given in terms of four real 
{\em gauge-invariant functions}.  In particular all quartic couplings are 
incorporated into one real, symmetric $4 \times 4$ matrix.  As we will show, 
in terms of the rge's of this quartic coupling matrix, symmetries of the 
THDM Higgs potential become very transparent.  Our key find is that there 
are cases in which two or three separate groups of terms have unequal 
symmetries and yet each retains its form even after renormalization.  We 
call this the phenomenon of {\em symbiotic symmetries}.\\

In order to make this article self-contained, we review here briefly the 
usage of {\em gauge-invariant functions}.  Consider first the 
most general potential of two Higgs doublets $\Phi_1$,$\Phi_2$ in the 
conventional notation~\cite{Gunion:1989we}
\begin{equation}
\label{Vconv}
\begin{split}
V =~& 
m_{11}^2 (\Phi_1^\dagger \Phi_1) +
m_{22}^2 (\Phi_2^\dagger \Phi_2) -
m_{12}^2 (\Phi_1^\dagger \Phi_2) -
(m_{12}^2)^* (\Phi_2^\dagger \Phi_1)\\
& +\frac{1}{2} \lambda_1 (\Phi_1^\dagger \Phi_1)^2 
+ \frac{1}{2} \lambda_2 (\Phi_2^\dagger \Phi_2)^2 
+ \lambda_3 (\Phi_1^\dagger \Phi_1)(\Phi_2^\dagger \Phi_2) \\ 
&+ \lambda_4 (\Phi_1^\dagger \Phi_2)(\Phi_2^\dagger \Phi_1)
+ \frac{1}{2} [\lambda_5 (\Phi_1^\dagger \Phi_2)^2 + \lambda_5^* 
(\Phi_2^\dagger \Phi_1)^2] \\ 
&+ [\lambda_6 (\Phi_1^\dagger \Phi_2) + \lambda_6^* 
(\Phi_2^\dagger \Phi_1)] (\Phi_1^\dagger \Phi_1) + [\lambda_7 (\Phi_1^\dagger 
\Phi_2) + \lambda_7^* (\Phi_2^\dagger \Phi_1)] (\Phi_2^\dagger \Phi_2).
\end{split}
\end{equation}
Hermiticity of the Lagrangian requires the parameters 
$m_{12}^2$, $\lambda_{5,6,7}$ to be complex and all other parameters to be real.
Owing to \ewgroup~gauge invariance, only terms of the form 
$(\Phi_i^\dagger \Phi_j)$ with $i,j=1,2$ may occur in the Higgs potential.
The Hermitian, positive semi-definite $2 \times 2$ matrix of all possible 
scalar products of this form may be decomposed in the following 
way~\cite{Maniatis:2006fs,Nishi:2006tg},
\begin{equation}
\label{eq-kmat}
\tmat{K} :=
\begin{pmatrix}
  \Phi_1^{\dagger}\Phi_1 & \Phi_2^{\dagger}\Phi_1 \\
  \Phi_1^{\dagger}\Phi_2 & \Phi_2^{\dagger}\Phi_2
\end{pmatrix}
 = \frac{1}{2} \left( K_0 \unitmatrix_2 + K_i  \sigma_i \right),
% 
% \unitmatrix_2
%      + \tvec{K}\,\tvec{\sigma} \right)
\end{equation}
with Pauli matrices $\sigma_i$, $i=1,2,3$, and the convention of summing over 
repeated indices is adopted.  Specifically, these real gauge-invariant 
functions are defined as
\begin{equation}
\label{eqKphi}
%\begin{split}
K_0 = \Phi_1^\dagger \Phi_1 + \Phi_2^\dagger \Phi_2, ~~~
K_1 = \Phi_1^\dagger \Phi_2 + \Phi_2^\dagger \Phi_1, ~~~
K_2 = i\Phi_2^\dagger \Phi_1 - i\Phi_1^\dagger \Phi_2, ~~~ 
K_3 = \Phi_1^\dagger \Phi_1 - \Phi_2^\dagger \Phi_2. 
%\end{split}
\end{equation}
The matrix $\tmat{K}$ in~\eqref{eq-kmat} is positive semi-definite with 
two conditions for the gauge-invariant functions:
\begin{equation}
K_0 \ge 0, \qquad K_\alpha K_\alpha \equiv K_0^2 - K_1^2 - K_2^2 - K_3^2 \ge 0.
\end{equation}
For convenience, we introduce the shorthand vector notation 
$\tvec{K}=(K_1,K_2,K_3)^\trans$.  For any $K_0$ and $\tvec{K}$, it is possible 
to find doublet fields $\Phi_{1,2}$ obeying~\eqref{eqKphi}.  These doublets 
then form an gauge orbit.  In terms of the gauge-invariant functions, the 
general THDM potential may be written in the simple form
\begin{equation}
\label{VK}
V = V_2 + V_4,\qquad \text{with }
V_2 = \xi_{\alpha}  K_\alpha, \quad
V_4 = \eta_{\alpha \beta} K_\alpha K_\beta,
\end{equation}
where $\xi_\alpha$ is a real 4-vector and
 $\eta_{\alpha \beta}$ is a real, symmetric $4 \times 4$ matrix.
 Expressed in terms of the conventional parameters, these tensors
 read
\begin{equation}
\xi_\alpha=\frac{1}{2}
\begin{pmatrix}
m_{11}^2+m_{22}^2, & 
- 2 \re(m_{12}^2), &
 2 \im(m_{12}^2), &
 m_{11}^2-m_{22}^2
\end{pmatrix}
\end{equation}
and
\begin{equation}
\label{VK4}
\eta_{\alpha \beta} = \frac{1}{4}
\begin{pmatrix}
\frac{1}{2}(\lambda_1 + \lambda_2) + \lambda_3 & 
\re(\lambda_6+\lambda_7) & 
-\im(\lambda_6+\lambda_7) & 
\frac{1}{2}(\lambda_1 - \lambda_2) \\ 
\re(\lambda_6+\lambda_7) & 
\lambda_4 + \re(\lambda_5) & 
-\im(\lambda_5) & \re(\lambda_6-\lambda_7) 
\\ 
-\im(\lambda_6+\lambda_7) & 
-\im(\lambda_5) & \lambda_4 - \re(\lambda_5) & 
-\im(\lambda_6-\lambda_7) \\ 
\frac{1}{2}(\lambda_1 - \lambda_2) & 
\re(\lambda_6-\lambda_7) & 
-\im(\lambda_6 -\lambda_7) & 
\frac{1}{2}(\lambda_1 + \lambda_2) - \lambda_3
\end{pmatrix}.
\end{equation}
It was shown that the formalism of gauge-invariant functions is advantageous 
in describing THDM's. That is, conditions for stability, stationarity, 
electroweak symmetry breaking, and CP violation of any THDM Higgs potential 
are easily described.  Here we will show that this formalism also gives 
insight into the symmetries of the THDM.  We are especially interested in 
symmetries which are not violated by the rge's.  To this aim let us start 
with a translation of the rge's of the couplings $\lambda_{1,2,3,4,5,6,7}$ 
in the conventional notation of the potential~\eqref{Vconv} to the rge's 
of the parameters $\eta_{\alpha \beta}$.  The one-loop renormalization group 
equations for $\lambda_{1,2,3,4,5,6,7}$ (including the $U(1)_Y$ and $SU(2)_L$ 
gauge interactions with couplings $g_1$ and $g_2$, respectively) 
are given by~\cite{Cheng:1973nv,Ferreira:2009jb,Haber:1993an}:
\begin{align}
\label{8}
8 \pi^2 \frac{d \lambda_1}{dt} =& 6 \lambda_1^2 + 2 \lambda_3^2 
+ 2 \lambda_3 \lambda_4 + \lambda_4^2 + |\lambda_5|^2 + 12 |\lambda_6|^2  
\\ 
&- \lambda_1 \left( \frac{3}{2} g_1^2 + \frac{9}{2} g_2^2 \right) 
+ \frac{3}{8} g_1^4 + \frac{3}{4} g_1^2 g_2^2 + \frac{9}{8} g_2^4, 
\nonumber \\ 
\label{9}
8 \pi^2 \frac{d \lambda_2}{dt} =& 6 \lambda_2^2 + 2 \lambda_3^2 
+ 2 \lambda_3 \lambda_4 + \lambda_4^2 + |\lambda_5|^2 + 12 |\lambda_7|^2 
\\ 
& - \lambda_2 \left( \frac{3}{2} g_1^2 + \frac{9}{2} g_2^2 \right) 
+ \frac{3}{8} g_1^4 + \frac{3}{4} g_1^2 g_2^2 + \frac{9}{8} g_2^4, 
\nonumber \\ 
\label{10}
8 \pi^2 \frac{d \lambda_3}{dt} =& (\lambda_1 + \lambda_2)(3 \lambda_3 
+ \lambda_4) + 2 \lambda_3^2 + \lambda_4^2 + |\lambda_5|^2 + 2 |\lambda_6|^2 
+ 2 |\lambda_7|^2 + 4 \lambda_6 \lambda_7^* + 4 \lambda_6^* \lambda_7 
\\ 
& - \lambda_3 \left( \frac{3}{2} g_1^2 + \frac{9}{2} g_2^2 \right) 
+ \frac{3}{8} g_1^4 - \frac{3}{4} g_1^2 g_2^2 + \frac{9}{8} g_2^4, 
\nonumber \\ 
\label{11}
8 \pi^2 \frac{d \lambda_4}{dt} =& (\lambda_1 + \lambda_2)\lambda_4 
+ 4 \lambda_3 \lambda_4 + 2 \lambda_4^2 + 4 |\lambda_5|^2 + 5 |\lambda_6|^2 
+ 5 |\lambda_7|^2 + \lambda_6 \lambda_7^* + \lambda_6^* \lambda_7 
\\ 
& - \lambda_4 \left( \frac{3}{2} g_1^2 + \frac{9}{2} g_2^2 \right) 
+ \frac{3}{2} g_1^2 g_2^2, 
\nonumber \\
\label{12} 
8 \pi^2 \frac{d \lambda_5}{dt} =& \lambda_5  \left( \lambda_1 + \lambda_2 
+ 4 \lambda_3 + 6 \lambda_4 \right) + 5 \lambda_6^2 + 5 \lambda_7^2 
+ 2 \lambda_6 \lambda_7 
\\ 
\nonumber
&- \lambda_5 \left( \frac{3}{2} g_1^2 + 
\frac{9}{2} g_2^2 \right), 
\\  
\label{13} 
8 \pi^2 \frac{d \lambda_6}{dt} =&  6 \lambda_1  \lambda_6 + 3 \lambda_3 
(\lambda_6 + \lambda_7) + \lambda_4 (4 \lambda_6 + 2 \lambda_7) + \lambda_5 
(5 \lambda_6^* + \lambda_7^*)
\\ 
&- \lambda_6 \left( \frac{3}{2} g_1^2 + \frac{9}{2} g_2^2 \right), 
\nonumber \\
\label{14}
8 \pi^2 \frac{d \lambda_7}{dt} =&  6 \lambda_2  \lambda_7 + 3 \lambda_3 
(\lambda_6 + \lambda_7) + \lambda_4 (2 \lambda_6 + 4 \lambda_7) + \lambda_5 
(\lambda_6^* + 5 \lambda_7^*) 
\\ 
&- \lambda_7 
\left( \frac{3}{2} g_1^2 + \frac{9}{2} g_2^2 \right).   
\nonumber 
\end{align}
In terms of $\eta_{\alpha \beta}$, they become
\begin{align}
\label{15}
8 \pi^2 \frac{d \eta_{00}}{dt} =& 4 \eta_{00}^2 + \eta_{00} (\eta_{11} 
+ \eta_{22} + \eta_{33}) + \eta_{11}^2 + \eta_{22}^2 + \eta_{33}^2 
+ 6 (\eta_{01}^2 + \eta_{02}^2 + \eta_{03}^2) 
\\ 
\nonumber
&+ 2 (\eta_{12}^2 + \eta_{13}^2 + \eta_{23}^2) - \eta_{00} \left( \frac{3}{2} 
g_1^2 + \frac{9}{2} g_2^2 \right) + \frac{3}{4} g_1^4 + \frac{9}{4} g_2^4, 
\\ 
\label{16}
8 \pi^2 \frac{d \eta_{01}}{dt} =& \eta_{01} \left( 6 \eta_{00}  
- \frac{3}{2} g_1^2 - \frac{9}{2} g_2^2 \right) + 6 ( \eta_{01} \eta_{11} 
+ \eta_{02} \eta_{12} + \eta_{03} \eta_{13}),
\\
\label{17}
8 \pi^2 \frac{d \eta_{02}}{dt} =& \eta_{02} \left( 6 \eta_{00}  
- \frac{3}{2} g_1^2 - \frac{9}{2} g_2^2 \right) + 6 ( \eta_{01} \eta_{12} 
+ \eta_{02} \eta_{22} + \eta_{03} \eta_{23}),
\\
\label{18}
8 \pi^2 \frac{d \eta_{03}}{dt} =& \eta_{03} \left( 6 \eta_{00}  
- \frac{3}{2} g_1^2 - \frac{9}{2} g_2^2 \right) + 6 ( \eta_{01} \eta_{13} 
+ \eta_{02} \eta_{23} + \eta_{03} \eta_{33}),
\\
\label{19}
8 \pi^2 \frac{d \eta_{11}}{dt} =& \eta_{11} \left( 3 \eta_{00} + 3 \eta_{11} - 
\eta_{22} - \eta_{33} - \frac{3}{2} g_1^2 - \frac{9}{2} g_2^2 \right) + 
\frac{3}{2} g_1^2 g_2^2 
\\ 
\nonumber
&+ 6 \eta_{01}^2 + 4 (\eta_{12}^2 + \eta_{13}^2),
\\
\label{20}
8 \pi^2 \frac{d \eta_{22}}{dt} =& \eta_{22} \left( 3 \eta_{00} - \eta_{11} + 
3 \eta_{22} - \eta_{33} - \frac{3}{2} g_1^2 - \frac{9}{2} g_2^2 \right) + 
\frac{3}{2} g_1^2 g_2^2 
\\
\nonumber
&+ 6 \eta_{02}^2 + 4 (\eta_{12}^2 + \eta_{23}^2),
\\
\label{21}
8 \pi^2 \frac{d \eta_{33}}{dt} =& \eta_{33} \left( 3 \eta_{00} - \eta_{11} - 
\eta_{22} + 3 \eta_{33} - \frac{3}{2} g_1^2 - \frac{9}{2} g_2^2 \right) 
+ \frac{3}{2} g_1^2 g_2^2 
\\ 
\nonumber
&+ 6 \eta_{03}^2 + 4 (\eta_{13}^2 + \eta_{23}^2), 
\\
\label{22}
8 \pi^2 \frac{d \eta_{12}}{dt} =& \eta_{12} \left( 3 \eta_{00} + 3 \eta_{11} 
+ 3 \eta_{22} - \eta_{33} - \frac{3}{2} g_1^2 - \frac{9}{2} g_2^2 \right) 
+ 6 \eta_{01} \eta_{02} + 4 \eta_{13} \eta_{23},
\\
\label{23}
8 \pi^2 \frac{d \eta_{13}}{dt} =& \eta_{13} \left( 3 \eta_{00} + 3 \eta_{11} 
- \eta_{22} + 3 \eta_{33} - \frac{3}{2} g_1^2 - \frac{9}{2} g_2^2 \right) 
+ 6 \eta_{01} \eta_{03} + 4 \eta_{12} \eta_{23},
\\
\label{24}
8 \pi^2 \frac{d \eta_{23}}{dt} =& \eta_{23} \left( 3 \eta_{00} - \eta_{11} 
+ 3 \eta_{22} + 3 \eta_{33} - \frac{3}{2} g_1^2 - \frac{9}{2} g_2^2 \right) 
+ 6 \eta_{02} \eta_{03} + 4 \eta_{12} \eta_{13}.
\end{align}
We now look for symmetries among the couplings $\eta_{\alpha \beta}$ which 
are preserved by the rge's.  Whereas~\eqref{8} to \eqref{14} are not 
particularly illuminating, \eqref{15} to \eqref{24} tell us immediately 
that the three conditions
\begin{equation}
\label{sp1}
\eta_{01} = \eta_{02} = \eta_{03}, \quad \eta_{11} = \eta_{22} = \eta_{33}, 
\quad \eta_{12} = \eta_{13} = \eta_{23},
\end{equation}
are preserved by the rge's.
Even though the rge's in general mix the quartic couplings, these 
conditions are maintained by them.  The corresponding quartic part of 
the potential is given by
\begin{equation}
\label{V4symb}
V_4^{\text{symb 5)}} = \eta_{00} K_0^2 + \eta_{11} (K_1^2 + K_2^2 + K_3^2) 
+ 2 \eta_{12} (K_1 K_2 + K_1 K_3 + K_2 K_3)
+ 2 \eta_{01} K_0 (K_1 + K_2 + K_3).
\end{equation}
With respect to the classification which we introduce later, this quartic 
part of the Higgs potential is denoted as case 5).  This quartic potential 
has the apparent symmetry $S_3$, generated by $K_1 \to K_2 \to K_3 \to K_1$ 
and $K_1 \to K_2 \to K_1$.  If $\eta_{01} = 0$, denoted as case 9), then the 
symmetry is $S_3 \times Z_2$ from having in addition the transformation 
$K_{1,2,3} \to -K_{1,2,3}$.  If $\eta_{12} = 0$ as well, denoted as case 12), 
then the symmetry is $O(3)$.  These three unequal symmetries coexist in 
$V_4^{\text{symb 5)}}$ even after renormalization.  This is our first example 
of the phenomenon of {\em symbiotic symmetries}.  In terms of $\Phi_{1,2}$, the 
respective symmetries are $Z_6$ (generated by $e=\frac{1}{2} (\unitmatrix_2 
- i[\sigma_1 + \sigma_2 + \sigma_3])$), $Q_{12}$ (generated by $e$ and $c_3 
= (i/\sqrt{2})(\sigma_1-\sigma_2)$), and $SU(2)$.  Case 12) with the 
symmetry $SU(2)$ was already known more than 30 years 
ago~\cite{Deshpande:1977rw}.  We now recognize the other possible  
symmetries with nonzero $\lambda_{6,7}$ for the first time.  They are 
likely to remain hidden if not for the Pauli formalism.\\

The quadratic part of the Higgs potential complementing the quartic 
part~\eqref{V4symb} is given by
\begin{equation}
\label{V4symbV2}
V_2^{\text{symb 5)}}= \xi_0 K_0 + \xi_1 \left(K_1 + K_2 + K_3 \right)
\end{equation}
Note that the rge's of the couplings~$\eta_{\alpha \beta}$, \eqref{15} to 
\eqref{24}, depend only on the quartic parameters themselves and not on the 
quadratic $\xi_\alpha$ parameters. Therefore any THDM with the quartic part 
$V_4$ given by~\eqref{V4symb} and arbitrary quadratic parameters~$\xi_\alpha$ 
will maintain its symbiotic symmetries. In the conventional notation, the 
conditions~\eqref{sp1} read
\begin{equation}
\label{cond5}
\begin{split}
&\im m_{12}^2 = -\re m_{12}^2 = \frac{1}{2} (m_{11}^2 - m_{22}^2), \quad
2 \lambda_4 = \lambda_1+\lambda_2-2 \lambda_3,\quad
\re (\lambda_5) =0,\\
&\re (\lambda_7) - \re (\lambda_6) = \im (\lambda_5),\quad
\re (\lambda_7) + \re (\lambda_6) = \frac{1}{2}(\lambda_1 - \lambda_2),\quad
\im (\lambda_6) = -\re (\lambda_6),\quad
\im (\lambda_7) = -\re (\lambda_7),
\end{split}
\end{equation}
so that this potential with all its symmetries is of the form
\begin{equation}
\label{Vconvsymb}
\begin{split}
V^{\text{symb 5)}} =& 
\phantom{+} m_{11}^2 \bigg[ (\Phi_1^\dagger \Phi_1) + \re (\Phi_1^\dagger \Phi_2)
+ \im (\Phi_1^\dagger \Phi_2)\bigg]
\\
&+ m_{22}^2 \bigg[ (\Phi_2^\dagger \Phi_2) - \re (\Phi_1^\dagger \Phi_2)- \im 
(\Phi_1^\dagger \Phi_2)\bigg]
\\
& +\frac{1}{2} \lambda_1 (\Phi_1^\dagger \Phi_1)^2 
+ \frac{1}{2} \lambda_2 (\Phi_2^\dagger \Phi_2)^2 
+ \lambda_3 (\Phi_1^\dagger \Phi_1)(\Phi_2^\dagger \Phi_2) \\ 
&+ \frac{1}{2}(\lambda_1+\lambda_2 -2 \lambda_3) (\Phi_1^\dagger \Phi_2)
(\Phi_2^\dagger \Phi_1)
- \im(\lambda_5) \im(\Phi_1^\dagger \Phi_2)^2\\
&+ \bigg[\frac{1}{2} (\lambda_1-\lambda_2) - \im (\lambda_5) \bigg] 
(\Phi_1^\dagger \Phi_1)
[\re(\Phi_1^\dagger \Phi_2) + \im(\Phi_1^\dagger \Phi_2)]\\
&+ \bigg[\frac{1}{2} (\lambda_1-\lambda_2) + \im (\lambda_5) \bigg] 
(\Phi_2^\dagger \Phi_2)
[\re(\Phi_1^\dagger \Phi_2) + \im(\Phi_1^\dagger \Phi_2)].
\end{split}
\end{equation}
As it appears, the underlying symmetries of this potential are far from 
being obvious.  We have thus demonstrated the utility of the Pauli formalism, 
and its use in all future studies of the THDM is advised.\\

Note that this potential is CP conserving. This follows from the sufficient
condition that all parameters in the potential are real, or it can be inferred 
from the necessary and sufficient conditions in~\cite{Maniatis:2007vn}. 
Hence this model has five Higgs bosons with definite CP properties. 
There are two charged Higgs bosons $H^\pm$, two CP even Higgs bosons $h^0$ 
and $H^0$, as well as one CP odd pseudoscalar Higgs boson $A^0$. The 
conditions for stability and electroweak symmetry breaking 
$\ewgroup \to U(1)_{\text{em}}$ are derived in a straightforward manner 
using the methods described in~\cite{Maniatis:2006fs}.  Here, stability 
in the {\em strong} sense (i.e. guaranteed by the quartic terms) requires
both conditions~\eqref{Vcond1} and \eqref{Vcond2} to be fulfilled:
\begin{equation}
\label{Vcond1}
\eta_{00}+\eta_{11}+ 2 \eta_{12} > 2 \sqrt{3} \big|\eta_{01}\big|\\
\end{equation}
\begin{equation}
\label{Vcond2}
\frac{3 \eta_{01}^2}{(\eta_{11} + 2 \eta_{12})^2} \ge 1\quad
\text{ or }\quad
\frac{3 \eta_{01}^2}{\eta_{11} + 2 \eta_{12}} < \eta_{00}
\end{equation}
For the model to have spontaneous symmetry breaking 
\mbox{$\ewgroup \rightarrow U(1)_{\text{em}}$}, we also require 
the condition
\begin{equation}
\xi_0 < \sqrt{3} \big| \xi_1 \big|.
\end{equation}
In the same way, the global minimum of the Higgs potential can be analytically 
obtained among the stationary solutions.  Under the assumption that we have 
chosen parameter values such that the potential is stable and has the required
electroweak symmetry breaking behavior, the masses of the neutral physical 
Higgs bosons are
\begin{equation}
\begin{split}
%&m_{H^\pm}^2 = \frac{2 v^2}{\xi_0^2-3 \xi_1^2} 
%\left\{
%6 \xi_0 \xi_1 \eta_{01}  - \xi_0^2(\eta_{11}+2\eta_{12})-3 \xi_1^2 \eta_{00}
%+ \sqrt{3} 
%\big| \xi_0 \xi_1 (\eta_{00}+ \eta_{11} +2\eta_{12}) - \eta_{01}(\xi_0^2+3 
%\xi_1^2) \big| \right\},\\
&m_{A^0} = m_{H^\pm}^2 +2 v^2 ( \eta_{11} - \eta_{12}),\\
&m_{h/H}^2 = 
\frac{1}{2} m_{H^\pm}^2 + v^2 (\eta_{11}+ \eta_{12})- \xi_0 - \xi_1\\
&\qquad \qquad \mp \sqrt{\frac{1}{4}(m_{H^\pm}^2
+2(v^2 (\eta_{11}+\eta_{12})+\xi_0))^2+\xi_1 (m_{H^\pm}^2+2 (v^2 (\eta_{11}
+\eta_{12})+\xi_0))+9 \xi_1^2}
\end{split}
\end{equation}
with $v \simeq$~246~GeV, being the SM vacuum expectation value.
The charged Higgs-boson mass $m_{H^\pm}$ follows directly
from the stationarity conditions. \\

Now let us consider more details regarding~\eqref{sp1}.  Examination 
of~\eqref{16} to \eqref{18} shows that the additional condition 
$\eta_{01} = \eta_{02} = \eta_{03}=0$ is also preserved by the rge's. 
This corresponds to 
\begin{equation}
\label{cond9}
\lambda_1 = \lambda_2
\end{equation}
in addition to the conditions~\eqref{cond5}.
%\begin{equation}
%\label{sp1a}
%\lambda_1 = \lambda_2 = \lambda_3 + \lambda_4, \quad \re(\lambda_5) = 0, \quad
%\im(\lambda_5) = -2 \re(\lambda_6) = 2 \im(\lambda_6), \quad \lambda_7 = -\lambda_6.
%\end{equation}
Using~\eqref{8} and \eqref{9} with \eqref{cond5}, we find
\begin{equation}
\label{27}
%8 \pi^2 \frac{d}{dt}(\lambda_1 - \lambda_3 - \lambda_4) = 
%(\lambda_1 - \lambda_3 - \lambda_4) \left( 6 \lambda_1 + 2 \lambda_4 - 
%\frac{3}{2} g_1^2 - \frac{9}{2} g_2^2 \right),
8 \pi^2 \frac{d}{dt}(\lambda_1 - \lambda_2) = 
(\lambda_1 - \lambda_2) \left( 6 \lambda_1 + 6 \lambda_2 -
12 \im(\lambda_5) 
-\frac{3}{2} g_1^2 - \frac{9}{2} g_2^2 \right),
\end{equation}
showing that indeed $\lambda_1=\lambda_2$ is a solution,
i.e. preserved by the rge's, even in the presence of the gauge 
couplings $g_{1,2}$.  In addition, \eqref{12} to \eqref{14} reduce to 
just one equation, i.e.
\begin{equation}
\label{28}
%8 \pi^2 \frac{d \im(\lambda_5)}{dt} = \im(\lambda_5) \left( 6 \lambda_3 + 
%8 \lambda_4 - 4 \im(\lambda_5) - \frac{3}{2} g_1^2 - \frac{9}{2} g_2^2 \right).
8 \pi^2 \frac{d \im(\lambda_5)}{dt} = \im(\lambda_5) \left( 8 \lambda_1 - 
2 \lambda_3 - 6 \im(\lambda_5) - \frac{3}{2} g_1^2 - \frac{9}{2} g_2^2 \right).
\end{equation}
As pointed out already, the resulting symmetry is $Q_{12}$ with character 
table given below.
\begin{table}[htb]
\begin{center}
\begin{tabular}{c|c|c|c|c|c|c|c}
$n$  & $h$ & $1^{++}$ & $1^{+-}$ & $1^{-+}$ & $1^{--}$ & 
$2^+$ & $2^-$ \\ 
\hline
1 & 1 & 1 & 1 & 1 & 1 & 2 & 2 \\ 
1 & 2 & 1 & --1 & --1 & 1 & 2 & --2 \\ 
2 & 3 & 1 & 1 & 1 & 1 & --1 & --1 \\ 
2 & 6 & 1 & --1 & --1 & 1 & --1 & 1 \\ 
3 & 4 & 1 & 1 & --1 & --1 & 0 & 0 \\ 
3 & 4 & 1 & --1 & 1 & --1 & 0 & 0 
\end{tabular}
\caption{Character table of $Q_{12}$.}
\end{center}
\end{table}

Examination of~\eqref{22} to \eqref{24} shows that the additional conditions
$\eta_{01} = \eta_{02} = \eta_{03}=0$, $\eta_{12} = \eta_{13} = \eta_{23}=0$
are also preserved by the rge's.  This case corresponds to 
$\im(\lambda_5) = 0$ in addition to~\eqref{cond5} and~\eqref{cond9}. 
It is obviously supported by~\eqref{28}, as discussed already 
in~\cite{Deshpande:1977rw}.  On the other hand, we see from~\eqref{19} 
to~\eqref{21} that the additional condition $\eta_{11} = \eta_{22} = 
\eta_{33}=0$ is {\em not} preserved by the rge's, which would have resulted 
in the symmetry $O(8)$.  Altogether we have found three related models, i.e.
the symbiotic model corresponding to the conditions~\eqref{sp1} followed 
by the models with the additional conditions $\eta_{01} =0$ and 
$\eta_{01} = \eta_{12} =0$, respectively.  These models, i.e. cases 5), 9), 
and (12), belong to one class of symbiotic models, denoted by class I) below.
\\

We now present the complete set of models with a symmetry consistent with the 
transparent rge's~\eqref{15} to \eqref{24}.  The symmetries found 
are summarized in Table~\ref{cases}.  In this table, the cases 5), 9), and 12) 
are discussed already.  In the columns 2 to 10 the conditions among the 
couplings $\eta_{\alpha \beta}$ are given. In the column denoted by `invariant 
terms', the allowed potential terms respecting the conditions are shown 
explicitly.  The last two columns then give the symmetries of the potential 
in addition to \ewgroup.  The next-to-last column gives the symmetry 
in terms of the $(\Phi_1, \Phi_2)^\trans$ basis. The cases 6) to 9) have
a quaternion symmetry in this basis.  The quaternion groups $Q_{4n}$ have 
$4n$ elements.  For example, $Q_8$ consists of $\pm 1, \pm a_{1,2,3}$ or 
$\pm 1, \pm a_3, \pm b_3, \pm c_3$, and $Q_{12}$ consists of $\pm 1, 
\pm c_{1,2,3}, \pm e, \pm e^2$ or $\pm 1, \pm b_{1,2}, \pm c_3, \pm d_3, 
\pm d_3^2$.  The last column gives the symmetry in terms of the $\tvec{K}$ 
basis.  The cases 5), 6), 9), and 12) have no variations.  All other cases 
have 3 variations each. Thus from the table we see that we have in total 28  
models.  In the $\Phi_{1,2}$ basis, there are several 
basic transformations which correspond to $K_i \to \pm K_j$.  Note that 
$\Phi_1 \to \Phi_2^\dagger$ (which would have meant $K_{1,3} \to K_{1,3}$, 
$K_2 \to -K_2$) is not allowed, and similarly for any transformation with 
determinant $-1$. We have suppressed such cases.

\newpage
\begin{center}
\begin{table}[htb]
%\caption{Special conditions on the $\eta$ parameters.}
\begin{tabular}{c|c|c|c||c|c|c||c|c|c|m{3.3cm}|m{3cm}|m{4cm}}
case  & $\eta_{01}$ & $\eta_{02}$ & $\eta_{03}$ & $\eta_{12}$ & $\eta_{13}$ & 
$\eta_{23}$ & $\eta_{11}$ & $\eta_{22}$ & $\eta_{33}$ & invariant terms & 
\mbox{symmetries $\!(\Phi_1,\Phi_2)^\trans$} & symmetries $\tvec{K}$ \\ 
\hline
1) & 0 & 0 & $\surd$ & $\surd$ & 0 & 0 & $\surd$ & $\surd$ & $\surd$ & 
\mbox{$K_3, K_1K_2, K_1^2$}, \mbox{$K_2^2, K_3^2$} & 
$a_3 = i \sigma_3 \in Z_4$ & 
\mbox{$A_3= \begin{pmatrix} -1 & 0 & 0 \\ 0 & \!\! -1 & 0 \\ 0 & 0 & 1\end{pmatrix}  
\in Z_2$}\\ [20pt]
2) & $\surd$ & $\eta_{01}$ & 0 & $\surd$ & $\surd$ & $-\eta_{13}$ & $\surd$ & 
$\eta_{11}$ & $\surd$ & 
\mbox{$K_1+K_2, K_1K_2$}, \mbox{$(K_1- K_2)K_3$},\newline \mbox{$K_1^2+K_2^2$}, 
\mbox{$K_3^2$} & 
$b_3 = \frac{i}{\sqrt{2}}(\sigma_1 + \sigma_2)
\newline 
\in Z_4$ & 
$B_3 =\begin{pmatrix} 0 & 1 & 0 \\ 1 & 0 & 0 \\ 0 & 0 & -1\end{pmatrix} 
\in Z_2$ \\ [20pt]
3) & $\surd$ & $-\eta_{01}$ & 0 & $\surd$ & $\surd$ & $\eta_{13}$ & $\surd$ & 
$\eta_{11}$ & $\surd$ & 
\mbox{$K_1-K_2$, $K_1K_2$},\newline \mbox{$(K_1+K_2)K_3$}, \mbox{$K_3^2$},
\newline \mbox{$K_1^2+K_2^2$} & 
$c_3 = \frac{i}{\sqrt{2}} (\sigma_1 - \sigma_2)
\newline
 \in Z_4$ &  
$ C_3=\begin{pmatrix} 0 &\!\! -1 & 0 \\ \!\! -1 & 0 & 0 \\ 0 & 0 &\!\! -1\end{pmatrix} 
\in Z_2$ \\ [20pt]
4) & $\surd$ & $\eta_{01}$ & $-\eta_{01}$ & $\surd$ & $-\eta_{12}$ & 
$-\eta_{12}$ & $\surd$ & $\eta_{11}$ & $\eta_{11}$ & 
\mbox{$K_1+K_2-K_3$}, \mbox{$K_1K_2 - (K_1+K_2)K_3$}, \mbox{$K_1^2+K_2^2+K_3^2$} &
$d_3\! =\! 
\newline
\frac{1}{2} (1\! +\! i[\sigma_1\! +\! \sigma_2\! -\! \sigma_3])$ 
\newline 
$\in Z_6$ & 
$D_3=\begin{pmatrix}0 & 0 & -1 \\ 1 & 0 & 0 \\ 0 & -1 & 0\end{pmatrix}  
\in S_3$ \\ [20pt]
5) & $\surd$ & $\eta_{01}$ & $\eta_{01}$ & $\surd$ & $\eta_{12}$ & $\eta_{12}$ 
& $\surd$ & $\eta_{11}$ & $\eta_{11}$ & 
\mbox{$K_1+K_2+K_3$}, \newline \mbox{$K_1K_2 + K_1K_3+K_2K_3$}, \newline 
\mbox{$K_1^2+K_2^2+K_3^2$} &
$e =
\newline
\frac{1}{2} (1 - i[\sigma_1 + \sigma_2 + \sigma_3])$
\newline 
$\in Z_6$ & 
$E=\begin{pmatrix}0 & 0 & 1 \\ 1 & 0 & 0 \\ 0 & 1 & 0\end{pmatrix} 
\in S_3$ \\ 
6) & 0 & 0 & 0 & 0 & 0 & 0 & $\surd$ & $\surd$ & $\surd$ & 
\mbox{$K_1^2, K_2^2, K_3^2$} &
$a_{1,2} = i\sigma_{1,2} \in Q_8$ & 
$\begin{matrix}
\phantom{0}\\
A_{1,2} \in Z_2 \times Z_2 \times Z_2
\\ \phantom{0}
\end{matrix}$
\\
7) & 0 & 0 & 0 & $\surd$ & 0 & 0 & $\surd$ & $\eta_{11}$ & $\surd$ &
\mbox{$K_1K_2$}, \mbox{$K_1^2+K_2^2$}, $K_3^2$ &
$a_3, b_3 \in Q_8$ & $A_3, B_3 \in  Z_2 \times Z_2$ \\
8) & 0 & 0 & 0 & $\surd$ & $-\eta_{12}$ & $-\eta_{12}$ & $\surd$ & $\eta_{11}$ 
& $\eta_{11}$ &
\mbox{$K_1K_2-(K_1+K_2)K_3$},\newline \mbox{$K_1^2+K_2^2+K_3^2$} &
$d_3, b_1 \in Q_{12}$ 
& 
$\begin{matrix}
\phantom{0}\\
D_3, B_1 \in  S_3 \times Z_2
\\ \phantom{0}
\end{matrix}$
\\
9) & 0 & 0 & 0 & $\surd$ & $\eta_{12}$ & $\eta_{12}$ 
& $\surd$ & $\eta_{11}$ & $\eta_{11}$ &
\mbox{$K_1K_2 + K_1K_3+K_2K_3$},\newline \mbox{$K_1^2+K_2^2+K_3^2$} &
$e, c_3 \in Q_{12}$ & 
$\begin{matrix}
\phantom{0}\\
E, C_3 \in  S_3 \times Z_2
\\ \phantom{0}
\end{matrix}$
\\
10) & 0 & 0 & $\surd$ & 0 & 0 & 0 & $\surd$ & $\eta_{11}$ & $\surd$ & 
\mbox{$K_3, K_1^2 + K_2^2, K_3^2$} &
$r_3(\theta)\! =\! 
\newline 
\cos(\frac{\theta}{2})\! +\! i \sin(\frac{\theta}{2}) 
\sigma_3$\newline $\in U(1)$ & 
\mbox{$R_3(\theta)\!=\!\begin{pmatrix} c_\theta &\!\! -s_\theta & 0 \\ s_\theta  
& c_\theta & 0 \\ 0 & 0 & 1\end{pmatrix}\! \in \!\! O(2)$} \\
11) & 0 & 0 & 0 & 0 & 0 & 0 & $\surd$ & $\eta_{11}$ & $\surd$ & 
\mbox{$K_1^2 + K_2^2, K_3^2$} &
$r_3(\theta), b_3 \in U(1) \times Z_2$ & 
$\begin{matrix}
\phantom{0}\\
R_3(\theta), B_3 \in O(2) \times Z_2
\\ \phantom{0}
\end{matrix}$
\\
12) & 0 & 0 & 0 & 0 & 0 & 0 & $\surd$ & $\eta_{11}$ & $\eta_{11}$ & 
\mbox{$K_1^2 + K_2^2 + K_3^2 $} &
$SU(2)$ & $O(3)$ 
\end{tabular}
\caption{\label{cases}\em Symmetries preserved by the rge's. In 
the first nine columns following the case number, the conditions with 
respect to the couplings~$\eta_{\alpha \beta}$ are given. Explicitly, the 
corresponding invariant terms of the THDM Higgs potential are shown. The 
next-to-last column gives the corresponding symmetry of the potential 
in the $(\Phi_1, \Phi_2)^\trans$ basis, whereas the last column gives the 
symmetry in terms of the $\tvec{K}$ basis. In case 10) in the last column 
we use the abbreviations $c_\theta = \cos(\theta)$ and $s_\theta = 
\sin(\theta)$. }
\end{table}
\end{center}

We may also classify the various cases of symbiotic symmetries. This 
was discussed already for the case 5) where we have subcases 9) and 12) 
with additional conditions. In this sense, we group the related models 5), 
9), and 12) in one class of models. In an analogous way we find the 
classes of related cases as given in Table~\ref{classes}. In addition, 
we give in this table for each class the symmetry of the models with 
respect to the $(\Phi_1, \Phi_2)^\trans$ basis.  Here, we have assumed that 
this basis is determined outside the Higgs potential, i.e. 
by their Yukawa couplings, which will of course break the symmetries we 
have discussed in this paper. However, except for the couplings proportional 
to $m_t$, they are small compared to the gauge couplings.  If all Yukawa 
couplings are neglected, then class II) in the above is equivalent to 
class I) because $b_3^\dagger e b_3 = d_3^\dagger$, and class V) is 
equivalent to class IV) because $a_1^\dagger b_3 a_1 = c_3$.

\newpage
\begin{center}
\begin{table}[htb]
\begin{tabular}{m{1cm}|m{3cm}|l}
class & related cases & symmetries $(\Phi_1, \Phi_2)^\trans$ \\
\hline
I)  & $5) \to 9) \to 12)$: & $Z_6 \to Q_{12} \to SU(2)$ \\
II) & $4) \to 8) \to 12)$: & $Z_6 \to Q_{12} \to SU(2)$ \quad [3 variations] \\ 
III)& $10) \to 11)$: & $U(1) \to U(1) \times Z_2$ \quad [3 variations] \\ 
IV) & $2) \to 7) \to 11)$: & $Z_4 \to Q_8 \to U(1) \times Z_2$ \quad 
[3 variations] \\ 
V)  & $3) \to 7) \to 11)$: & $Z_4 \to Q_8 \to U(1) \times Z_2$ \quad 
[3 variations] \\ 
VI) & $1) \to 6)$: & $Z_4 \to Q_8$ \quad [3 variations]
\end{tabular}
\caption{\label{classes} \em 
Classes of symbiotic symmetries for all related cases shown 
in Table~\ref{cases}.  The subsequent cases originate from the previous 
case by an additional condition. Also given are the symmetries with 
respect to the $(\Phi_1, \Phi_2)^\trans$ basis and the number of variations.}
\end{table}
\end{center}

Let us summarize our findings.  Recently it was shown that by using Pauli 
matrices, the formalism of gauge-invariant functions simplifies the study 
of THDM's.  We have determined the renormalization-group equations of the 
parameters $\eta_{\alpha \beta}$ of the Higgs potential in this approach.  
In so doing, relations among these couplings become completely transparent, 
allowing us to find all possible symmetries which are preserved by the rge's. 
We discover cases where the quartic Higgs potential has two 
or more separate sectors with unequal symmetries, but are nevertheless 
maintained by the rge's, including the gauge interactions.  We call this 
the phenomenon of {\em symbiotic symmetries}. In a systematic way, we have 
obtained all possible models with a symmetry beyond that of the SM, as shown 
in Table~\ref{cases}.  There are 12 basic scenarios, 8 of which have 
3 variations, for a total of 28 such models. There are 6 symbiotic classes 
as shown in Table~\ref{classes}.  These symmetries are very much hidden 
in the $\lambda_{1,2,3,4,5,6,7}$ parameterization of the Higgs potential, 
but become totally transparent in the Pauli formalism.\\

{\em Acknowledgments}~:~ We thank the organizers (especially Maria Krawczyk 
and Ilya Ginzburg) of the Workshop on Nonminimal Higgs Models at Lake Baikal 
(July 2009) for affording us the opportunity to embark on this project.  
The work of E.M. was supported in part by the U.~S.~Department of Energy 
under Grant No.~DE-FG03-94ER40837.\\

%%%%%%%%%%%%%%%%%%%%%%%%%%%%%%%%%%%%%%%%%%%%%%%%%%%%%%%%%%%%%%%%%%%%%%%
%
% REFERENCES
%
%%%%%%%%%%%%%%%%%%%%%%%%%%%%%%%%%%%%%%%%%%%%%%%%%%%%%%%%%%%%%%%%%%%%%%%

\bibliographystyle{unsrt}

\begin{thebibliography}{99}


%%%%%%%%%%%%%%%%%%%%%%%%%%%%%%%%%%%%%%%%%%%%%%%%%%%%%%%%%%%%%%%%%
% THDM
%%%%%%%%%%%%%%%%%%%%%%%%%%%%%%%%%%%%%%%%%%%%%%%%%%%%%%%%%%%%%%%%%

%\cite{Lee:1973iz}
\bibitem{Lee:1973iz}
  T.~D.~Lee,
  %``A Theory of Spontaneous T Violation,''
  Phys.\ Rev.\  D {\bf 8} (1973) 1226.
  %%CITATION = PHRVA,D8,1226;%%

%%%%%%%%%%%%%%%%%%%%%%%%%%%%%%%%%%%%%%%%%%%%%%%%%%%%%%%%%%%%%%%%%
% THDM different aspects
%%%%%%%%%%%%%%%%%%%%%%%%%%%%%%%%%%%%%%%%%%%%%%%%%%%%%%%%%%%%%%%%%

%\cite{Kobayashi:1973fv}
\bibitem{Kobayashi:1973fv}
  M.~Kobayashi and T.~Maskawa,
  %``CP Violation In The Renormalizable Theory Of Weak Interaction,''
  Prog.\ Theor.\ Phys.\  {\bf 49}, 652 (1973).
  %%CITATION = PTPKA,49,652;%%

%%%%%%%%%%%%%%%%%%%%%%%%%%%%%%%%%%%%%%%%%%%%%%%%%%%%%%%%%%%%%%%%%
% Ma's initial article 
%%%%%%%%%%%%%%%%%%%%%%%%%%%%%%%%%%%%%%%%%%%%%%%%%%%%%%%%%%%%%%%%%

%\cite{Deshpande:1977rw}
\bibitem{Deshpande:1977rw}
  N.~G.~Deshpande and E.~Ma,
  %``Pattern Of Symmetry Breaking With Two Higgs Doublets,''
  Phys.\ Rev.\  D {\bf 18}, 2574 (1978).
  %%CITATION = PHRVA,D18,2574;%%

%\cite{Gunion:1989we}
\bibitem{Gunion:1989we}
  J.~F.~Gunion, H.~E.~Haber, G.~L.~Kane and S.~Dawson,
  ``THE HIGGS HUNTER'S GUIDE.''
  %%CITATION = BNL-41644;%%

%\cite{Cvetic:1993cy}
\bibitem{Cvetic:1993cy}
  G.~Cvetic,
  %``CP violation in bosonic sector of SM with two Higgs doublets,''
  Phys.\ Rev.\  D {\bf 48}, 5280 (1993)
  [arXiv:hep-ph/9309202].
  %%CITATION = PHRVA,D48,5280;%%

%%%%%%%%%%%%%%%%%%%%%%%%%%%%%%%%%%%%%%%%%%%%%%%%%%%%%%%%%%%%%%%%%
% Nagel's thesis
%%%%%%%%%%%%%%%%%%%%%%%%%%%%%%%%%%%%%%%%%%%%%%%%%%%%%%%%%%%%%%%%%

%\cite{Nagel:2004sw}
\bibitem{Nagel:2004sw}
  F.~Nagel, PhD thesis (Heidelberg, 2004) 
  ``New aspects of gauge-boson couplings and the Higgs sector''
% \href{/spires/find/hep/www?irn=6461018}{SPIRES entry}

   %\cite{Ginzburg:2004vp}
\bibitem{Ginzburg:2004vp}
  I.~F.~Ginzburg and M.~Krawczyk,
  %``Symmetries of two Higgs doublet model and CP violation,''
  Phys.\ Rev.\  D {\bf 72}, 115013 (2005)
  [arXiv:hep-ph/0408011].
  %%CITATION = PHRVA,D72,115013;%%
  
  %\cite{Gunion:2005ja}
\bibitem{Gunion:2005ja}
  J.~F.~Gunion and H.~E.~Haber,
  %``Conditions for CP-violation in the general two-Higgs-doublet model,''
  Phys.\ Rev.\  D {\bf 72}, 095002 (2005)
  [arXiv:hep-ph/0506227].
  %%CITATION = PHRVA,D72,095002;%%
  
%\cite{Branco:2005em}
\bibitem{Branco:2005em}
  G.~C.~Branco, M.~N.~Rebelo and J.~I.~Silva-Marcos,
  %``CP-odd invariants in models with several Higgs doublets,''
  Phys.\ Lett.\  B {\bf 614}, 187 (2005)
  [arXiv:hep-ph/0502118].
  %%CITATION = PHLTA,B614,187;%%

\bibitem{Ma:2006km}
  E.~Ma, 
  Phys.\ Rev.\  D {\bf 73}, 077301 (2006)
  [arXiv:hep-ph/0601225].

%\cite{Nishi:2006tg}
\bibitem{Nishi:2006tg}
  C.~C.~Nishi,
  %``CP violation conditions in N-Higgs-doublet potentials,''
  Phys.\ Rev.\  D {\bf 74}, 036003 (2006)
  [Erratum-ibid.\  D {\bf 76}, 119901 (2007)]
  [arXiv:hep-ph/0605153].
  %%CITATION = PHRVA,D74,036003;%%
  
%\cite{Maniatis:2006fs}
\bibitem{Maniatis:2006fs}
  M.~Maniatis, A.~von Manteuffel, O.~Nachtmann and F.~Nagel,
  %``Stability and symmetry breaking in the general two-Higgs-doublet model,''
  Eur.\ Phys.\ J.\  C {\bf 48}, 805 (2006)
  [arXiv:hep-ph/0605184].
  %%CITATION = EPHJA,C48,805;%%
  
%\cite{Fromme:2006cm}
\bibitem{Fromme:2006cm}
  L.~Fromme, S.~J.~Huber and M.~Seniuch,
  %``Baryogenesis in the two-Higgs doublet model,''
  JHEP {\bf 0611}, 038 (2006)
  [arXiv:hep-ph/0605242].
  %%CITATION = JHEPA,0611,038;%%
  
%\cite{Barroso:2007rr}
\bibitem{Barroso:2007rr}
  A.~Barroso, P.~M.~Ferreira and R.~Santos,
  %``Neutral minima in two-Higgs doublet models,''
  Phys.\ Lett.\  B {\bf 652}, 181 (2007)
  [arXiv:hep-ph/0702098].
  %%CITATION = PHLTA,B652,181;%%
  
%\cite{Gerard:2007kn}
\bibitem{Gerard:2007kn}
  J.~M.~Gerard and M.~Herquet,
  %``A twisted custodial symmetry in the two-Higgs-doublet model,''
  Phys.\ Rev.\ Lett.\  {\bf 98}, 251802 (2007)
  [arXiv:hep-ph/0703051].
  %%CITATION = PRLTA,98,251802;%%


%%%%%%%%%%%%%%%%%%%%%%%%%%%%%%%%%%%%%%%%%%%%%%%%%%%%%%%%%%%%%%%%%
% THDM in Nachtmann formalism
%%%%%%%%%%%%%%%%%%%%%%%%%%%%%%%%%%%%%%%%%%%%%%%%%%%%%%%%%%%%%%%%%


%\cite{Maniatis:2007vn}
\bibitem{Maniatis:2007vn}
  M.~Maniatis, A.~von Manteuffel and O.~Nachtmann,
  %``CP Violation in the General Two-Higgs-Doublet Model: a Geometric View,''
  Eur.\ Phys.\ J.\  C {\bf 57} (2008) 719
  [arXiv:0707.3344 [hep-ph]].
  %%CITATION = EPHJA,C57,719;%%


%\cite{Nishi:2007dv}
\bibitem{Nishi:2007dv}
  C.~C.~Nishi,
  %``Physical parameters and basis transformations in the Two-Higgs-Doublet
  %model,''
  Phys.\ Rev.\  D {\bf 77}, 055009 (2008)
  [arXiv:0712.4260 [hep-ph]].
  %%CITATION = PHRVA,D77,055009;%%


  
%%%%%%%%%%%%%%%%%%%%%%%%%%%%%%%%%%%%%%%%%%%%%%%%%%%%%%%%%%%%%%%%%
% Generalized CP transformations in the THDM
%%%%%%%%%%%%%%%%%%%%%%%%%%%%%%%%%%%%%%%%%%%%%%%%%%%%%%%%%%%%%%%%%

%\cite{Ferreira:2009wh}
\bibitem{Ferreira:2009wh}
  P.~M.~Ferreira, H.~E.~Haber and J.~P.~Silva,
  %``Generalized CP symmetries and special regions of parameter space in the
  %two-Higgs-doublet model,''
  Phys.\ Rev.\  D {\bf 79}, 116004 (2009)
  [arXiv:0902.1537 [hep-ph]].
  %%CITATION = PHRVA,D79,116004;%%



%%%%%%%%%%%%%%%%%%%%%%%%%%%%%%%%%%%%%%%%%%%%%%%%%%%%%%%%%%%%%%%%%
% Stability and perturbativity of the couplings in the THDM
%%%%%%%%%%%%%%%%%%%%%%%%%%%%%%%%%%%%%%%%%%%%%%%%%%%%%%%%%%%%%%%%%

%\cite{Ferreira:2009jb}
\bibitem{Ferreira:2009jb}
  P.~M.~Ferreira and D.~R.~T.~Jones,
  %``Bounds on scalar masses in two Higgs doublet models,''
  JHEP {\bf 0908} (2009) 069
  [arXiv:0903.2856 [hep-ph]].
  %%CITATION = JHEPA,0908,069;%%


%\cite{Mahmoudi:2009zx}
\bibitem{Mahmoudi:2009zx}
  F.~Mahmoudi and O.~Stal,
  %``Flavor constraints on the two-Higgs-doublet model with general Yukawa
  %couplings,''
  [arXiv:0907.1791 [hep-ph]].
  %%CITATION = ARXIV:0907.1791;%%


%%%%%%%%%%%%%%%%%%%%%%%%%%%%%%%%%%%%%%%%%%%%%%%%%%%%%%%%%%%%%%%%%
% Fayet or Peccei-Quinn symmetry
%%%%%%%%%%%%%%%%%%%%%%%%%%%%%%%%%%%%%%%%%%%%%%%%%%%%%%%%%%%%%%%%%

%\cite{Fayet:1974pd}
\bibitem{Fayet:1974pd}
  P.~Fayet,
  %``Supergauge Invariant Extension Of The Higgs Mechanism And A Model For The
  %Electron And Its Neutrino,''
  Nucl.\ Phys.\  B {\bf 90}, 104 (1975).
  %%CITATION = NUPHA,B90,104;%%



%%%%%%%%%%%%%%%%%%%%%%%%%%%%%%%%%%%%%%%%%%%%%%%%%%%%%%%%%%%%%%%%%
% absense of FCNC by imposing Z_2
%%%%%%%%%%%%%%%%%%%%%%%%%%%%%%%%%%%%%%%%%%%%%%%%%%%%%%%%%%%%%%%%%

%\cite{Glashow:1976nt}
\bibitem{Glashow:1976nt}
  S.~L.~Glashow and S.~Weinberg,
  %``Natural Conservation Laws For Neutral Currents,''
  Phys.\ Rev.\  D {\bf 15}, 1958 (1977).
  %%CITATION = PHRVA,D15,1958;%%
           

%\cite{Paschos:1976ay}
\bibitem{Paschos:1976ay}
  E.~A.~Paschos,
  %``Diagonal Neutral Currents,''
  Phys.\ Rev.\  D {\bf 15}, 1966 (1977).
  %%CITATION = PHRVA,D15,1966;%%


%\cite{Peccei:1977ur}
\bibitem{Peccei:1977ur}
  R.~D.~Peccei and H.~R.~Quinn,
  %``Constraints Imposed By CP Conservation In The Presence Of Instantons,''
  Phys.\ Rev.\  D {\bf 16}, 1791 (1977).
  %%CITATION = PHRVA,D16,1791;%%
  
%\cite{Peccei:1977hh}
\bibitem{Peccei:1977hh}
  R.~D.~Peccei and H.~R.~Quinn,
  %``CP Conservation In The Presence Of Instantons,''
  Phys.\ Rev.\ Lett.\  {\bf 38}, 1440 (1977).
  %%CITATION = PRLTA,38,1440;%% 
  




%%%%%%%%%%%%%%%%%%%%%%%%%%%%%%%%%%%%%%%%%%%%%%%%%%%%%%%%%%%%%%%%%
% Ma, symmetries
%%%%%%%%%%%%%%%%%%%%%%%%%%%%%%%%%%%%%%%%%%%%%%%%%%%%%%%%%%%%%%%%%

%\cite{Ma:1977td}
\bibitem{Ma:1977td}
  E.~Ma,
  %``Symmetry Of The Higgs Potential In SU(N) Gauge Models,''
  Nucl.\ Phys.\  B {\bf 132}, 317 (1978).
  %%CITATION = NUPHA,B132,317;%%


%%%%%%%%%%%%%%%%%%%%%%%%%%%%%%%%%%%%%%%%%%%%%%%%%%%%%%%%%%%%%%%%%
% CEL
%%%%%%%%%%%%%%%%%%%%%%%%%%%%%%%%%%%%%%%%%%%%%%%%%%%%%%%%%%%%%%%%%

%\cite{Cheng:1973nv}
\bibitem{Cheng:1973nv}
  T.~P.~Cheng, E.~Eichten and L.~F.~Li,
  %``Higgs Phenomena In Asymptotically Free Gauge Theories,''
  Phys.\ Rev.\  D {\bf 9}, 2259 (1974).
  %%CITATION = PHRVA,D9,2259;%%


%%%%%%%%%%%%%%%%%%%%%%%%%%%%%%%%%%%%%%%%%%%%%%%%%%%%%%%%%%%%%%%%%
% beta-functions in the minimal susy model
%%%%%%%%%%%%%%%%%%%%%%%%%%%%%%%%%%%%%%%%%%%%%%%%%%%%%%%%%%%%%%%%%

%\cite{Haber:1993an}
\bibitem{Haber:1993an}
  H.~E.~Haber and R.~Hempfling,
  %``The Renormalization group improved Higgs sector of the minimal
  %supersymmetric model,''
  Phys.\ Rev.\  D {\bf 48}, 4280 (1993)
  [arXiv:hep-ph/9307201].
  %%CITATION = PHRVA,D48,4280;%%






\end{thebibliography}

\end{document}